# RECURRENCE PLOT AND RECURRENCE QUANTIFICATION ANALYSIS TECHNIQUES FOR DETECTING A CRITICAL REGIME. EXAMPLES FROM FINANCIAL MARKET INDICES

## A. FABRETTI[1,2,], M. AUSLOOS[2,†]

[1] Department of Mathematics for Economy, Insurances and Finance Applications, University of Roma 1, La Sapienza I-00100 Rome, Italy
[2] SUPRATECS, B5, University of Liège, B-4000 Liège, Euroland

Keywords: endogenous crash, financial bubble, Recurrence Plot, Recurrence Quantification Analysis, nonlinear time series analysis, DAX, NASDAQ

Abstract. Recurrence Plot (RP) and Recurrence Quantification Analysis (RQA) are signal numerical analysis methodologies able to work with non linear dynamical systems and non stationarity. Moreover they well evidence changes in the states of a dynamical system. We recall their features and give practical recipes. It is shown that RP and RQA detect the critical regime in financial indices (in analogy with phase transition) before a bubble bursts, whence allowing to estimate the bubble initial time. The analysis is made on DAX and NASDAQ daily closing price between Jan. 1998 and Nov. 2003. DAX is studied in order to set-up overall considerations, and as a support for deducing technical rules. The NASDAQ bubble initial time has been estimated to be on Oct. 19, 1999.

## 1. Introduction

The occurrence of a financial crash is a tragic event for many investors. Recent papers have shown some analogy between crashes and phase transitions [1, 2, 3]. Moreover like in earthquakes, log periodic oscillations have been found before some crashes [4, 5]. It was then proposed that an economic index $y(t)$ increases as a complex power law, whose first order Fourier representation is

$$y(t) = A + B\ln(t_c - t)\{1 + C\cos[\omega\ln(t_c - t) + \phi]\} \tag{1.1}$$

where $A$, $B$, $C$, $\omega$, $\phi$ are constants and $t_c$ is the critical time (rupture time). From best fits with this formula it has been shown that one can predict financial crashes [4]. Notwithstanding this does not answer the question why a crash occurs. A crash is thought to occur when the system goes out of equilibrium and enters an unstable phase where any information is amplified [6]. It should be noted that a crash necessarily happens during a period of generalized economic euphoria, with high returns, high volatility, high correlation between markets and unprecedent imitation behavior by investors. This critical period takes the name of 'speculative bubble'.

A financial time series is often studied from the point of view of stochastic processes [7]. On the other hand it is well known that non linear simple deterministic systems can exhibit the phenomenon of deterministic chaos [8, 9]. The debate on the appropriate approach, i.e. if it has to be stochastic or instead must see these events from the point of view of deterministic chaos, generated by non linear dynamics, is quite open [10, 11]. A topological approach to analyze chaos developed in the physics literature [12, 13] based on close return plots, for example, has produced much evidence that financial data are not chaotic [14], where chaotic is meant as something unordered, unpredictable and confusing.

We have investigated the possibility to use Recurrence Plot (RP) and Recurrence Quantification Analysis (RQA) techniques for financial time series, in particular for detecting critical regimes preceding an endogenous crash. Recurrence Plots are graphical tools elaborated by Eckmann, Kamphorst and Ruelle in 1987 and are based on Phase Space Reconstruction [15]. In 1992, Zbilut and Webber [16] proposed a statistical quantification of RPs and gave it the name of 'Recurrence Quantification Analysis' (RQA). Here RP and RQA techniques are not intended to prove evidence of chaos, but are used for their goodness in working with non stationarity and noisy data [17], in detecting changes in data behavior, in particular in detecting breaks, like a phase transition [18], and in informing about other dynamic properties of a time series [15].

Most of the applications of RP and RQA are at this time in the field of physiology and biology as for example protein folding analysis [19], but some authors have already applied the techniques to financial data [10, 11, 14, 20, 21, 22, 23, 24, 25, 26, 27, 28] and cloud base height fluctuation [29]. In financial application some authors have chosen these techniques for their goodness to work with non linearity [20] or they applied to foreign exchange data [21] for no stationarity, others for testing chaos [22] or noise. Thiel et al. in [30] have tested the quantity of information contained in RP claiming that it is possible to reconstruct a time series from its unthresholded RP. Even if the choice of threshold and embedding parameters can condition the quantity of information that can be read in an RP, " .. this new framework provides a useful complementary tool to test for complexity in financial data.. " [20]. A complete bibliography on RP and RQA and their application can be found in the web page by Marwan [31], while a recent overview is found in [32].

The following work is an application of RP and RQA techniques in studying rare events, like crashes in financial markets and in particular to search whether these techniques can distinguish between normal and critical regime, in the thermodynamic sense, whence give an estimation of the initial bubble time. The analysis is here below made on two time series, NASDAQ and DAX, taken over a time span of 6 years including the known crash of April 2000 [5]. The series are also divided into subseries in order to investigate changes in the evolution of the signal.

The treatment is divided into two parts, a theoretical and a practical one. The first part is dedicated to recall the theoretical basis of 'Phase Space Reconstruction' (Sect. 2) with attention on the methods used to choose so called embedding parameters, to explain how to plot an RP (Sect. 3) and to define the RQA variables (Sect. 4). The formulas for each variable are found in Appendix A. Several examples of RPs and RQA on how to interpret the results are also found and discussed in Sect. 3 and 4. Moreover a simulation of an econophysically interesting log periodic signal generated from Eq. (1.1) has been made and is discussed in Sect. 4.

The second part is dedicated to our specific analysis. In Sect. 5 the data and the methodology analysis are presented. In Sect. 6 "practical considerations" are outlined about the choice of embedding parameters for financial signals. Analysis of DAX and NASDAQ series are discussed respectively in Sect. 7 and 8. In Sect. 9 conclusions are found.

## 2. Phase space reconstruction

Recurrence Plots are graphical tools based on Phase Space Reconstruction [15]. The changing state of a dynamic system can be indeed represented by sequences of 'state vectors' in the phase space, a so called 'phase-flow' [33]. Usually in a dynamic system a detailed analysis is possible when the equations of motion and all degrees of freedom $n$ are known. Unfortunately only a few quantities can be usually observed in a system. However from a theoretical point of view it is argued that it is possible to reconstruct the entire dynamics of a system from a relatively small number of observables; because the different degrees of freedom of a dynamic system interact with each other, the combination of all other components is concealed in each observable quantity through the main state vector components. Let $\xi_{t+1} = F(\xi_t)$ be a deterministic system, let $(X_1, ..., X_N)$ be $N$ observations of some variable of this system. According to Takens' theorem [34] it is possible to reconstruct a phase space trajectory from such a single scalar time series of observable quantities $(X_1, ..., X_N)$. In so doing it is possible to answer a wide range of questions about the system by examining the dynamics in a space defined by delayed vectors of dimension $m$. Construct a delayed vector $y(i) = X_i, X_{i+d}, ..., X_{i+(m-1)d}$, where $d$ and $m$ are called respectively the time delay and the embedding dimension, together they are called the embedding parameters. Therefore each unknown point of the phase space at time $i$ is reconstructed by the delayed vector $y(i)$ in an $m$-dimensional space called the reconstructed phase space.

2.1. **Embedding parameters.** The most natural question pertains on how to choose an appropriate value for the time delay $d$ and the embedding dimension $m$. Several methods have been developed to best guess $m$ and $d$. The most often used methods are the Average Mutual Information Function (AMI) for the time delay, as introduced by Fraser and Swinney in 1986 [35] and the False Nearest Neighbors (FNN) method for the embedding dimension developed by Kennel et al. [36].

2.1.1. *Time delay.* First of all, the time delay has to be estimated, since most logically the method to find the embedding dimension needs an estimation of $d$.

There are two main methods. In the first one, the value for which the autocorrelation function

$$C(d) = \frac{1}{N-d} \sum_{i=1}^{N-d} (X_i - \bar{X})(X_{i+d} - \bar{X}) \tag{2.1}$$

first passes through zero is searched, which gives $d$. In the second, one chooses the first minimum location of the average mutual information function, where the mutual information function is defined as follows. Let us start with partitioning the real numbers. Let $p_i$ be the probability to find a time series value in the $i - th$ interval of the partition, let $p_{ij}(d)$ be the joint probability to find a time series value in the $i - th$ interval and a time series value in the $j - th$ interval after a time $d$, i.e. the probability of transition in $d$ time from the $i - th$ to the $j - th$ interval. The average mutual information function is

$$S(d) = -\sum_{ij} p_{ij}(d) \ln \frac{p_{ij}(d)}{p_i p_j}. \tag{2.2}$$

The value $d$ that firstly minimizes the quantity $S(d)$ is the method choice for finding a reasonable time delay.

The difference between these two methods resides in the fact that while the first looks for linear independence, the second measures a general dependence of two variables. For this reason the second method seems to be preferred in non linear time series analysis.

2.1.2. *Embedding dimension.* The method used to find the embedding dimension is based on the concept of false neighbor. A false neighbor is a point in the data set that looks like a neighbor to another because the orbit is seen in a too small embedding space. For example two points on a circle can appear close to each other, even though they are not, if e.g. the circle is seen sideways (as a projection), thus is appearing like a line segment, whence.

increasing by one the dimension $m$ of the reconstructed space often permits to differentiate between the points of the orbit, i.e those which are true neighbors and those which are not.

Let $y(i)$ be a point of the reconstructed space. Note as $y(i)^r$ the $r - th$ nearest neighbor and compute the Euclidean distance $L_2$ between them

$$R_m^2(y(i), y(i)^r) = \sum_{k=0}^{m-1} [y(i+kd) - y^r(i+kd)]^2. \qquad (2.3)$$

Next increase $m$ to $m+1$ and compute the new distance, i.e. $R_{m+1}^2(y(i), y(i)^r)$

$$R_{m+1}^2(y(i), y(i)^r) = R_m^2(y(i), y(i)^r) + [y(i+md) - y^r(i+md)]^2. \qquad (2.4)$$

The point $y^r(i)$ is said a false nearest neighbor if

$$[\frac{R_{m+1}^2(y(i), y(i)^r) - R_m^2(y(i), y(i)^r)}{R_m^2(y(i), y(i)^r)}] > R_{tol} \qquad (2.5)$$

where $R_{tol}$ is a predefined threshold. Note that the number of false nearest neighbors depends on $R_{tol}$. The sensitivity of the criterion to $R_{tol}$ is not discussed here. Kennel et al. found that for $R_{tol} \geq 10$ the false nearest neighbor is clearly identified and we stick by this value below, but for a more profound discussion we suggest to read [36].

In practice, the percentage of false nearest neighbors (FNN) is computed for each $m$ of a set of values; the embedding dimension is said to be found for the first percentage of FNN dropping to zero. Notice that when the signal is noisy this percentage never reaches a true zero value.

For finding the embedding parameters the TISEAN version 2.1 code by Hegger and Schreiber [37] was used. TISEAN is a software package for nonlinear time series analysis freely available on the web [38].

The values of embedding parameters for the simulated data here below (Sect. 3.1) are given in Table 1. More practical considerations on the choice of embedding parameters from experimental data with some discussions are left for section 6.

## 3. RECURRENCE PLOT

The Recurrence Plot (RP) is a matrix of points $(i, j)$ where each point is said to be recurrent and marked with a dot if the distance between the delayed vectors $y(i)$ and $y(j)$ is less than a given threshold $\varepsilon$; the distance can be $L_1$, Euclidean or $L_\infty$. In general the threshold $\varepsilon$ has to be chosen as small as possible, but a too small $\varepsilon$ can lead one to miss some structure, if there is noise distortion. As each coordinate $i$ represents a point in time, RP provides information about the temporal correlation of phase space points [15]. Indeed each horizontal coordinate $i$ in RP refers to the state of the system at $i$ and each vertical coordinate $j$ refers to the state in $j$. So if the point $(i, j)$ is marked as recurrent, the state $j$ belongs to the neighborhood centered in $i$ of size $\varepsilon$; this means that the state of the system at time $i$ has some 'similarity' with the state of the system at $j$, in other words we can say that the system is staying on nearby 'orbits'.

The software used for RP and RQA is the Visual Recurrence Analysis (VRA) version 4.6 code by Kononov, freely available on the web [39]. We skipped the problem of the threshold using a scale of color for distances; on each plot, white 'color' represents a short distance while 'black' corresponds to a long distance as it can be seen in the legend on the right side of each plot, in which each 'color' indicates a range of distances between states.

3.1. **Typical cases.** For a stationary signal the dependence should be only on the distance between $i$ and $j$ but not on their position, so that an RP should appear homogeneous. Conversely homogeneity in RP indicates stationarity, as seen for example on Figure 1, showing a Brownian Motion and its RP. By extension if the texture is homogeneous within an RP block, a stationary behavior can be deduced for the corresponding subperiod.

Vertical or horizontal lines on an RP denote that the system state does not change or change very slowly in time. Diagonal lines correspond to trajectories passing in the same region of the phase space at different times. Therefore parallel and perpendicular lines to the main diagonal appear when the series presents some determinism or periodicity; see Figure 2 for a sinusoid and its RP. The length of lines parallel to the main diagonal of the RP indicates how fast the trajectories diverge in phase space.

In pure stochastic systems no parallel or vertical lines appear on an RP, as can be seen in Figure 3, LHS showing a white noise signal with its RP on the RHS.

In general an aggregation of points indicates persistent non stationarity; black bands or areas indicate rare or extreme events.

Therefore RPs can be used to test a system deterministic behavior through the percentage of recurrent points belonging to parallel lines. In so doing, RPs are useful tools for the preprocessing of experimental time series and provide a comprehensive image of the dynamic course at a glance [18]. In appendix A a summary of typical features can be found.

3.2. **A log-periodic simulation.** In order to study the crash from the point of view of a phase transition with log periodic precursors, i.e. equation (1.1), a log periodic signal with trend has been simulated. First the law (1.1) with $C = 0$

$$y_{trend}(t) = A + B \ln(t_c - t), \tag{3.1}$$

is studied; see Fig 4(LHS). Then the contribution of the log-periodicity generated by

$$y_{log-per}(t) = \cos[\omega \ln(t_c - t) + \phi], \tag{3.2}$$

is done; see Fig. 5(LHS). The simulation of the combination of both terms, i.e. the function (1.1), is shown in Fig. 6(LHS).

We can deduce a few interesting points from the displayed features. When a system is not stationary and has a strong trend an 'arrow' shape appears as can be seen in Figure 4(RHS),corresponding to the simulation of the function (3.1). Note the smooth border (color) lines limiting well defined areas on this plot.

In Fig 5(RHS) the RP of the mere log periodic oscillations (3.2) is plotted. As can been seen, as the period decreases, the white lines, indicating periodicity, as in Fig. 2(RHS), are not parallel to the main diagonal anymore but converge toward the upper right corner; this is because the log-periodic oscillations are more frequently occurring near the critical time.

The two contributions (3.1) and (3.2) together generate the RP plotted in Fig. 6. The 'arrow' shape is due to the trend, as in Fig. 4; the not too smooth border ('color') lines are due to the log periodicity, as seen in Fig 5.

At this time it could be interesting also to consider a phase transition signature. In Fig. 7(LHS) an arbitrary signal is plotted before and after a peak, whence near a crash, taking into account some overall shape like that of an anti-bubble after a crash [40]. The RP aspect of Fig. 7(RHS) reveals a feature far from the normal signal evolution, that is the crash time interval. Notice the well marked black bands corresponding to the crash time.

## 4. Recurrence Quantification Analysis

In 1992, Zbilut and Webber [16] proposed to quantify the presence of patterns, like parallel lines of RPs, through statistical values and gave it the name 'Recurrence Quantification

Analysis' (RQA). As RP this methodology is independent of limiting constraints, like the data set size, and does not require stationarity, linearity and any usual assumptions on the probability distribution of data. For these reasons RQA, as the RP, seems very useful for characterizing state changes.

Initially Zbilut and Webber introduced 5 variables to quantify changes; for detailed formulas see the Appendix B. The 5 basic RQA variables are briefly defined as follows with a brief comment on each one:

**%REC:** the percentage of recurrent points; recall that a point $(i, j)$ is recurrent if the distance between the vectors $y(i)$ and $y(j)$ is less than the threshold; in other words %REC is the ratio of the number of recurrent states measured with respect to all possible states.

**%DET:** the percentage of recurrent points forming line segments parallel to the main diagonal. The presence of these lines reveals the existence of a deterministic structure.

**MAXLINE:** the longest line segment measured parallel to the main diagonal. In Trulla et al. [41], it is claimed that this quantity is proportional to the inverse of the largest positive Lyapunov exponent. A periodic signal produces long line segments, while short lines indicate chaos.

**TREND:** the slope of line-of-best-fit through %REC as a function of the displacement from the main diagonal (excluding the last 10% range). This variable quantifies the drift and the non stationarity of the time series. For example, a flat slope indicates stationarity because in an homogeneous plot the quantity of recurrent points at the left and at the right of the central line is almost the same; instead when on the RHS this quantity is less than the quantity on the LHS the variable TREND assumes negative values; this means that the system is going far away from the state it had, i.e. it has a trend.

**ENT:** the Shannon entropy of the distribution of the length of line segments parallel to the main diagonal. The entropy gives a measure of how much information one needs in order to recover the system. A low entropy value indicates that few informations are needed to identify the system, in contrast, a high entropy indicates that much informations are required. The entropy is small when the length of the longest segment parallel to the diagonal is short and does not vary much. This has to be associated with information on determinism. A high entropy is typical of periodic behavior while low entropy indicates chaotic behavior.

All these variables provide information about different aspects of the plot and are in fact inter-correlated.

For RQA computation VRA [39] has been used. It permits to do the computation on selected time intervals of variable sizes. Two complementary studies have been made: in the first case, the RQA variables have been computed for the whole time series, in the other case, we divided the whole time series into subseries and computed the variables in each subinterval, called an epoch. Each epoch is 100 days long and regularly shifted by 10 days, in such a way that each epoch overlaps the next one by 90 days. If $N_e$ is the length of each epoch and $d_e$ the shift, the epoch $i$ corresponds in the time series to the days starting in $t = (i-1)d_e + 1$ and ending in $t = (i-1)d_e + N_e + 1$.

The first investigation has been devised in view of comparing global effects due to structures in subseries, while the computation for various epochs has been made to emphasize the changes in state inside the whole time series.

A notice on the displayed data is here in order. For conciseness, the results presented in table forms correspond to the variables computed on the whole time series, while the Figures correspond to the variables computed on epochs.

In Table 2 the values of RQA variables are given for the Brownian motion, the sinusoid and the white noise in order to illustrate the simple cases studied in the RP section, in a quantitative way, while the RQA variables are given for the simulated trend, the log-periodicity and function 1.1 in Table 3.

As can be seen from the tables, high values of %REC, %DET, MAXLINE and ENT are associated to deterministic processes, while high values of TREND are associated with a not stationary process having a strong trend.

In Fig. 8, RQA variables values computed on epochs are plotted for the simulated phase transition plotted in Fig 7.

All the variables, studied for epoch, take high values because the signal is (more) deterministic-like. However when in coincidence with the phase transition, see epoch 49 for %REC, ENT and TREND, and see epoch 51 for %DET, the RQA variables take their lowest values. All these epochs include the critical time. Instead, the MAXLINE assumes the lowest values in epochs 42 and 57. Notice that the first epoch (42) stops before the phase transition while the second (57) starts at the critical time.

## 5. Data and Methodology

The financial index data used below are the daily closing price of the German index DAX and the American technology index NASDAQ. The period spanned by the data is Jan. 1998 till Nov. 2003, giving a total of 1493 daily observations for DAX and 1482 daily observations for NASDAQ. They are plotted in Fig. 9(LHS) and 10(RHS) respectively. Both time series have been first transformed into the classical momentum divided by the maximum $X_{max}$ of the time series over the time window which is considered in the specific analysis:

$$X_i^{norm} = \frac{X_i - X_1}{X_{max}}. \qquad (5.1)$$

RPs have been plotted for each time series, RQA variables have been computed on different epochs, see Fig. 11 and 12. The data has been afterwards divided into subseries and several of these analyzed again; see Tables 4 and 5 for the corresponding definitions of such subsets and the RQA variable values.

## 6. Practical considerations

At this stage, before embarking into a discussion of the financial index results, practical considerations are to be made. It should be recalled that a search for $d$ and $m$ values must be made first. This is done, in TISEAN, through the function 'mutual' which computes the 'mutual information' while the function 'false_nearest' computes the percentage of false nearest neighbor. In Fig. 13, the values of the mutual information and the percentage of FNN for the DAX (top) and for the NASDAQ (bottom) are shown. On the left, the mutual information function is displayed when computed for each time lag $d$, while on the figure RHS the percentage of false nearest neighbors is shown for each dimension $m$. Recall that the optimum time delay $d$ is chosen as the one which minimizes the average mutual information of equation (2.2), while the embedding dimension $m$ is chosen as the value for which the percentage of false nearest neighbors drops to zero.

Even if it is difficult to see a marked minimum in both Fig. 13(LHS) (top and bottom), for both DAX and NASDAQ the value chosen for the time delay is hereby taken equal to 10; while the embedding dimension can be chosen equal to 5. These values are those used below to reconstruct the space even when the subseries are analyzed.

Consider that each delayed vector is constructed along a period of $(m-1)d$ days, given a set of $d$ and $m$ values, with this choice the state space is reconstructed over ca. 2 months. In so doing, each vector $y(i)$ of the so reconstructed space somewhat represents the evolution of the market during these 2 months, with a 2 week lag; this kind of phase space reconstruction can be expected to allow one to recognize changes at an appropriate time. Larger parameters values imply long time intervals and in fact induce less sensitivity to changes on a short time scale.

In the following, RP is further used to have a glance about the behavior of DAX and NASDAQ along the studied time span. The discussion will be such that after observing 'epochs' in which the signal could be assumed to be stationary and about finding rare events, both financial index time series are divided into subseries, themselves reconsidered with RP and RQA techniques for finer discussion.

## 7. DAX

The RP of DAX between Jan.05, 98 and Nov.23, 03 is shown in Fig 9(RHS). It can be first observed that this RP is not homogeneous. In fact stationarity in such financial time series can hardly be assumed to occur over a period of 6 years. Let us zoom inside the RP and observe how each band (black, white or grey) can be interpreted. Between coordinates 0 and 430 the RP looks quite white. For this reason it can be deduced that during the first two years of data no (rare) remarkable event occurred, except for the DAX maximum reached in July 98, corresponding to the small grey band in the lower left hand corner in Fig 9(RHS). The dark grey area with an 'L' shape encircled in a clear grey area that is centered around the point of coordinates (530,200) corresponds to the end of 1999 and the beginning of 2000, when the behavior of DAX changes and the index has an upward acceleration. This feature corresponds to the acceleration ending in March/April 2000. Thereafter a bearish period started, going on until the beginning of 2003. The dark grey area centered around coordinates (1300,520) represents the difference between the bearish period at the beginning of 2003 and the bullish one at March 2000.

The values of RQA variables computed on different epochs are shown in Figure 11 for the DAX. It can be seen that the highest absolute values are found in the epochs 38 to 43, corresponding to July 1999 and March 2000, where an acceleration took place. Other high value epochs are between 82 and 87 corresponding to March 2001 and September 2001 respectively, when the DAX has the deepest descend during the bearish period. Notice that all the RQA variables assume the lowest values in epochs 13 to 35 corresponding to the end of July 1998 and the end of November 1999 when the behavior looks like stationary.

Therefore in view of these observations the DAX time series was divided first into two subseries: one going from Jan. 05, 1998 to Mar. 08, 2000, i.e. the bullish period, the other from March 09, 2000 to Jan. 23, 2003 which represents the bearish period. Moreover it has to be observed that the white square area on the lower left hand corner in the RP is the 'image' of the Jul. 28, 1998 to Nov. 29, 1999 time corresponding in RQA exactly to the epochs having the lowest values. No evident trend is present in this period, whence the corresponding region of RP looks homogeneous. Therefore the signal in that period can be assumed to be rather stationary and studied separately.

Fig. 14 is the RP of the bullish period from Jan. 05, 1998 to March 08, 2000, 552 days; the large white area in the upper right corner is a stationary period, instead on the RP borders the grey areas becoming darker are due to the drift.

Fig. 15 is the RP of the bearish period from March 08, 2000 to Jan. 23, 2003, 728 days; in the first 350 days the price has a soft negative trend that corresponds to the white area and a fading grey. The grey area delimited between coordinates $x = 350/400$ corresponds to the deep minimum in DAX on Sept. 21, 2001. After this minimum the index rises a little bit but then again a negative trend follows to form an 'arrow' shape; the dark grey area on the lower right corner shows the index difference in value just after the crash and after almost 2 years.

Fig. 16 is the RP for a stationary signal period from Jul. 28, 1998 to Nov. 29, 1999, 341 days. The RP looks quite homogeneous, there is a small black band on the lower left corner, but the stationarity is quite proven. The RQA variable values for these time series are given in Table 4. It is worth to note that the period considered as stationary is that with the lowest values, as already seen in RQA results when the technique is applied to distinct epochs.

## 8. Nasdaq

Fig. 10(RHS) is the RP of NASDAQ between Jan. 05, 1998 and Nov. 21, 2003. Of interest is the dark grey vertical band surrounded by a lighter grey area. The grey area is delimited by horizontal coordinates $x = 452$ and $x = 690$, the dark grey one is delimited by horizontal coordinates $x = 504$ and $x = 566$.

The grey band corresponds to Oct. 19, 1999 to Sept. 27, 2000; the dark grey band corresponds to Jan. 03, 2000 to March 31, 2000. The dark grey area suggests that those days are 'far away' from all other days in a phase space sense. In fact all white regions represent recurrence, while the black or grey areas represent rare events (see discussion in Sect. 3). Note the same RP shape of the phase transition simulation in Fig. 7 and note also that the dates here above fall in the same time interval as the bubble and the subsequent crash.

The corresponding RQA variables for NASDAQ are plotted in Fig. 12: the highest absolute values are reached in epochs between 37 and 43, corresponding to the period between June 10, 1999 and Jan. 28, 2000; this period includes the bubble growth.

In order to have more insight, let us observe that in contrast to the RP of DAX only one event emerges strongly from the plot, i.e. the crash.

Thus it can be supposed that the initial bubble time occurs at $x = 452$ (Oct. 19,1999), i. e. the coordinate of the border line of the grey area in Fig 10 (RHS); let us call it $t^*$. We can thus deduce that on such a day the evolution of the system changes, i.e. the evolution passes from a normal regime to a critical regime, as at the Ginzburg temperature [42] in phase transition studies.

In order to further analyze whether and how the data changes, the series has been divided into subseries of 200 days shifted by 100 days for each analysis, such that each sub-series is overlapping during almost 5 months.

For the first subseries (I) starting in Jan. 1998 and ending in Oct. 1998 the RP in Figure 17 presents a small dark grey area at $x = 130$, when the NASDAQ reached prices around 2000 in July 1998. The remaining part of the RP looks homogeneous. Notice that RQA variables in Table 5 do not show any remarkable value.

The second subseries (II) goes from June 1998 to Feb. 1999: the RP in Figure 18 is quite homogeneous except for some isolated not recurrent points aligned along the main diagonal. The RQA variables in Table 5 are a little bit larger than for I.

The third subseries (III) is from Oct. 1998 to July 1999; the RP in Figure 19 looks quite homogeneous except for some regions corresponding to local peaks. The RQA variables are smaller than for I and II.

The fourth subseries (IV) goes from Feb. 1999 to Dec. 1999; the RP in Figure 20 shows the characteristic shape typical of a strong trend as studied and pointed out in Fig. 4(RHS) and 6(RHS). The trend starts to be significant in the middle of Oct. 1999. This indicates that the RP has changed indeed when the bubble has started. The RQA variables in Table 5 evidence in this period the highest values. It has to be underlined that this IV period does not include the crash time, but stops before the bubble bursts.

It is worth to emphasize the similarity between Figures 6 and 20: it indicates a pattern in IV similar to the simulation of the log-periodic function (1.1).

The fifth V subseries goes from July 1999 till the end of May 2000. The RP shown in Figure 21 indicates a grey area which starts near $t^*$, in the plot corresponding with $x = 48$. However the bubble is not so evidently seen in this RP, as it was for the time interval studied above.

It has been also tried to detrend the data by the function (3.1), see Fig. 22(LHS). The difference of the data and the fitted function produces an oscillating series whose RP is shown in Fig 22(RHS).

## 9. CONCLUSION

One of the main aims of this review was first of all to indicate how to use RP and RQA in order to distinguish normal regimes from critical regimes. The case of financial index data series was used as an illustration. The techniques were applied to DAX and NASDAQ between Jan. 1998 and Nov. 2003 encompassing the crash of April 2000, i.e. presenting a bullish period followed by a bearish one.

As seen the RP and RQA results evidence very well defined differences in the state and evolution of signals when comparing data in different epochs. Indeed from an RP one can argue about the stationarity and the behavior of the series when looking at the homogeneity and at the different 'color bands'. From the values of RQA variables it can be argued about the deterministic behavior of the signal. Indeed if the RP does not look homogeneous and if black or dark grey bands are present, it can be argued that a rare event occurred, as seen in the simulation in Fig. 7(RHS). Moreover at the same time the RQA variables, in particular %DET and ENT, take high values. In general the low values of these variables can be associated to stochasticity, while high values are associated with a deterministic behavior, as seen in Table 2. The inhomogeneity of RP and the high values of RQA variables reveal events that can be interpreted as a thermodynamic phase transition, if the homogeneity and low values are restored after this event.

Turning toward specific data analysis, DAX and NASDAQ, looking at RPs in Fig. 9 and 10 one fact is evident: the same 'rare' event has occurred at the same time (around coordinates 500) in both markets. However this event is better emphasized through the grey band in the RP of the NASDAQ than in the RP of the DAX. In fact in Fig. 10 the grey band representing this event is quite isolated, suggesting that after this event a normal regime was restored, i.e. like when a phase transition occurs, see again Fig. 7. In Fig. 9 the same event is not so obvious but it can be still seen as a strong fluctuation, because in the DAX the event is more like an extended drop than a real crash.

Each coordinate in RP is linked with the time series, so the border line of a grey or black band reveals the time when the data behavior starts to change like for a thermodynamic transition. As seen in the previous section the border line of the grey band around the black one was delimited by coordinates 452 and 690 that correspond to the Oct. 19, 1999 and Sept. 27, 2000 days. This is an a posteriori estimation, but through the analysis of the fourth subseries IV one can argue to have found the beginning of the bubble, and have an estimate, with some delay, on the initial moment before the bubble grows, that in our cases corresponds to Dec. 16, 1999. This delay consists in $(m-1)d$ days (ca. 120 days). The fourth subseries IV goes from March 16, 1999 to Dec. 28, 1999. It includes the initial moment and gives the range of time necessary for the estimation. Notice that the subseries does not contain the crash time, but yet the features warn about its occurrence.

It is worth also to note the similarity between Figures 6 and 20. Starting from the point of view of log-periodic oscillation, the log periodic model in (1.1) was simulated and the generated signal was studied by RP. The resulting plot was compared with the RP of the fourth NASDAQ time series. Some similarity can be seen in them, the arrow shape is due to a strong trend, see Fig. 4; it has been seen that a rather smooth trend leads to a ' rather smooth' plot, while the effect of the log periodicity leads to border color line curves with projections like in Fig. 6. Nobody can expect from such an experimental data to have a truly smooth trend, in fact Fig. 20 has no smooth border line. Some of those curve lines could be due to the log periodicity but the presence of noise does not allow us to affirm it. The same consideration holds also for Fig. 7(RHS) and 10(RHS), and also for Fig. 5(RHS) and 22(RHS).

In conclusion, the possibility to use the RP and RQA methods in order to detect a bubble regime and to find the initial bubble time (by analogy to the Ginzburg temperature [42]) has been investigated. It has been shown that, with some delay $((m-1)d$ days) as respect to the beginning but enough time before the crash (3 months in this particular case), such that a warning could be given, RP and RQA detect a difference in state and recognize the critical regime.

## 10. Acknowledgements

The authors thank COSTP10 for an STSM grant to AF and acknowledge discussion with Claire Gilmore, Charles Webber and Giulia Rotundo.

## APPENDIX A. INTERPRETING RP

The following scheme is a summary of what has been discussed in section 3

```
-------------------------------------------------------------------
Homogeneity                   Stationarity
Darkening from the upper left  Non stationarity:drift and trend
  corner to the lower right corner
Black or dark grey bands       Non stationarity:
                                 some system states are far from
                                 the medium evolution or  can
                                 correspond to a transition.
Periodic patterns              There are cycling evolutions of
                                 the system; the distance
                                 between periodic patterns is
                                 the period.
Isolated recurrent points      Rapid fluctuations of the system,
                                 which could be random; in fact
                                 white noise looks like a set of
                                 isolated random points.
Diagonal lines parallel to the The system evolution is similar
  main diagonal                  at different times; the system
                                 could be deterministic.
Vertical and horizontal lines  Some state of the system does
                                 not change or do so slowly.
-------------------------------------------------------------------
```

## APPENDIX B. RECURRENCE QUANTIFICATION VARIABLES

Let us define

$$R_{ij} = \begin{cases} 1 & \text{if } (i,j) \text{ is recurrent,} \\ 0 & \text{otherwise.} \end{cases} \tag{B.1}$$

and consider that

$$D_{ij} = \begin{cases} 1 & \text{if } (i,j) \text{ and } (i+1,j+1) \text{ or } (i-1,j-1) \text{ are recurrent ,} \\ 0 & \text{otherwise.} \end{cases} \tag{B.2}$$

In terms of $R_{ij}$ and $D_{ij}$ the RQA variables are easily written as follows [43].
The percentage of recurrent points in an RP is

$$\%REC = \frac{1}{N^2} \sum_{i,j=1}^{N} R_{ij}. \tag{B.3}$$

The percentage of recurrence points which form diagonal lines is

$$\%DET = \frac{\sum_{i,j=1}^{N} D_{ij}}{\sum_{i,j=1}^{N} R_{ij}}. \tag{B.4}$$

Let be $N_l$ the number of diagonal lines and let be $l_i$ the length of the $i-th$ diagonal line.
The length of the longest diagonal line is

$$MAXLINE = \max(l_i; i = 1, ..., N_l). \tag{B.5}$$

The Shannon entropy for a function $f(x)$ is

$$H(x) = -\sum_{x \in dom(X)} f(x) log f(x),$$  (B.6)

while for a distribution it is

$$H(p_1, ..., p_n) = -\sum_{i=1}^{n} p_i log(p_i).$$  (B.7)

Let us denote $p(l)$ the distribution of the diagonal line lengths, its Shannon entropy is

$$ENT = -\sum_{l=1}^{N_l} p(l) \ln p(l).$$  (B.8)

Let us define

$$RR_i = \sum_{j=1}^{\tilde{N}} R_{ij}$$  (B.9)

where $\tilde{N}$ is the number of recurrent points after rejecting 10% of the farthest away occurring points; in other words $RR_i$ is the number of recurrence points with respect to the state $i$. The formula for the trend obtained minimizing the least square error is

$$TREND = \frac{\sum_{i=1}^{\tilde{N}} (i - \tilde{N}/2)(RR_i - \overline{RR_i})}{\sum_{i=1}^{\tilde{N}} (i - \tilde{N}/2)^2}$$  (B.10)

where $\overline{RR_i}$ denotes the average of $RR_i$ over $i \in [1, \tilde{N}]$.

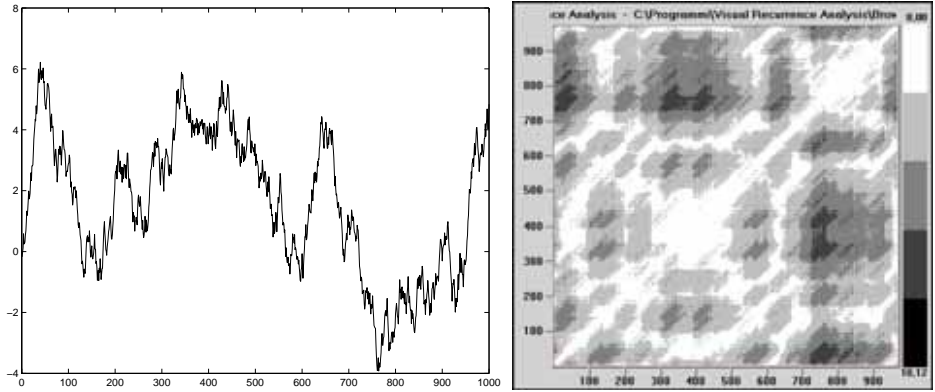

FIGURE 1. The figure on the right hand side (RHS) is the RP of the Brownian Motion for 1000 points plotted on the left hand side (LHS). The homogeneity of the RHS plot reveals stationarity

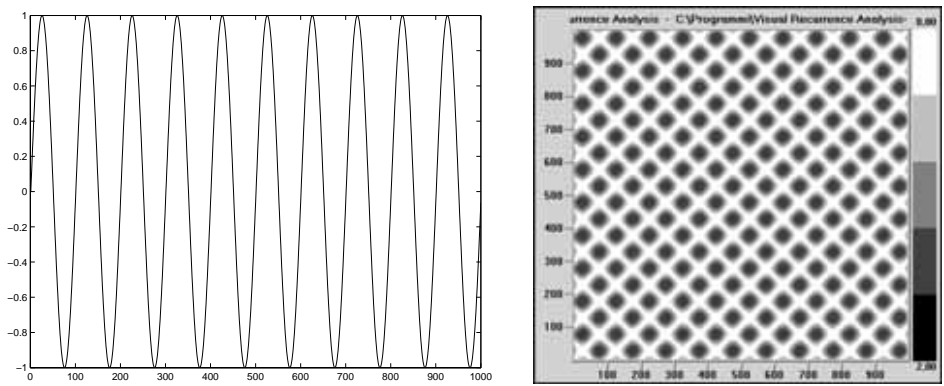

FIGURE 2. (LHS) a sin wave of 1000 points, (RHS) the corresponding RP. See the diagonal lines parallel and vertical to the main diagonal indicating that the signal is periodic. This kind of structure also indicates determinism

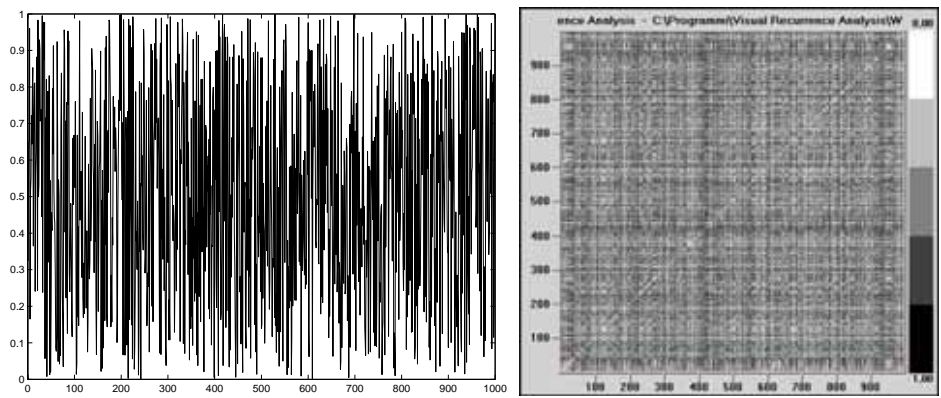

Figure 3. (RHS) the RP of a white noise signal, sample of 1000 points, plotted on the LHS. The RHS plot is homogeneous, without any remarkable pattern; this suggests stationarity and casuality.

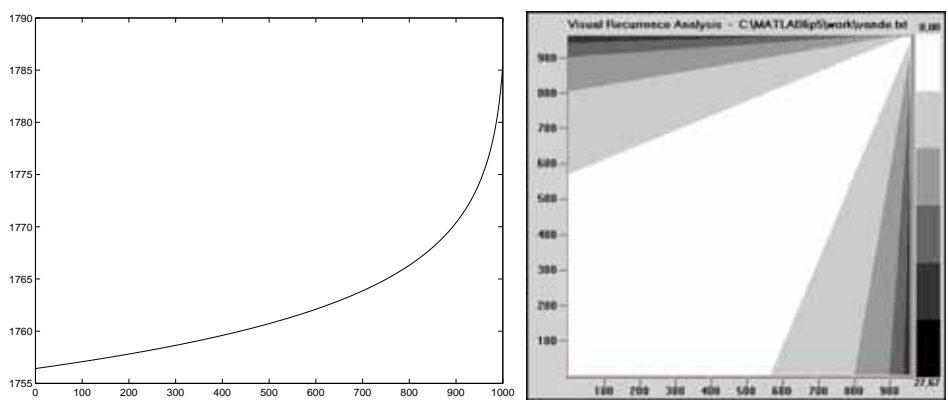

Figure 4. (RHS)RP of a signal on the LHS as generated by equation (3.1). The arrow shape on the RHS plot is the sign of a strong trend.

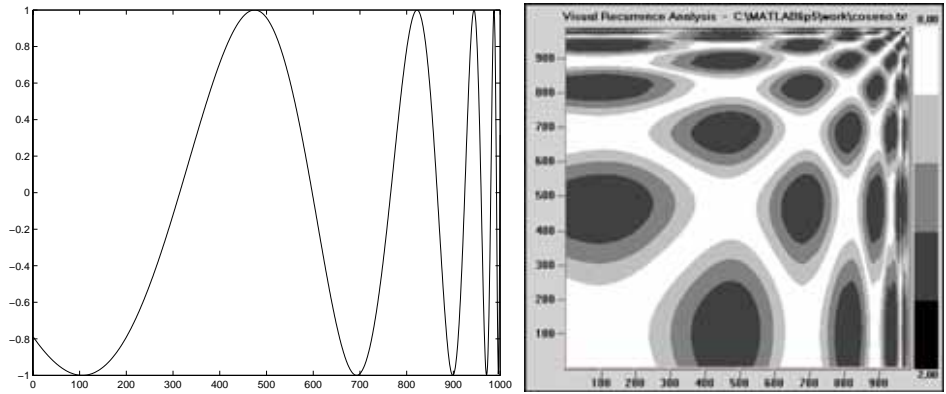

FIGURE 5. (RHS)RP of a log periodic signal as generated by equation (3.2), as shown on the LHS.

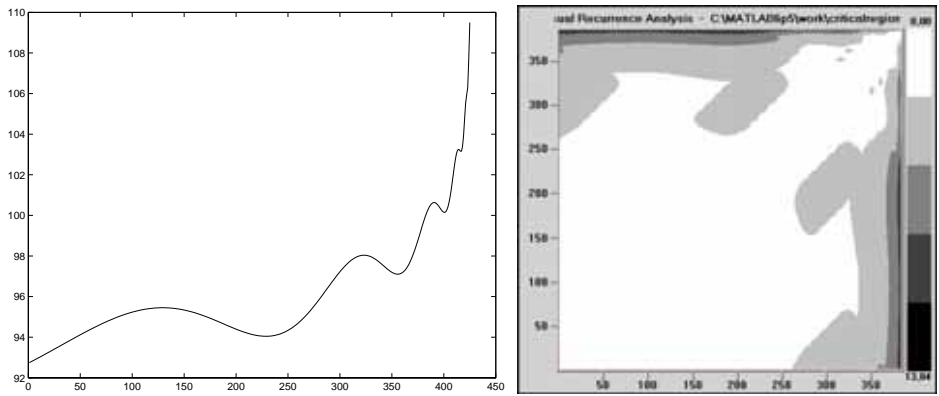

FIGURE 6. (RHS) RP of a log periodic signal as generated by equation (1.1), as shown on the LHS. The arrow shape on the RHS plot is the sign of a strong trend; the curve lines are due to the log periodicity.

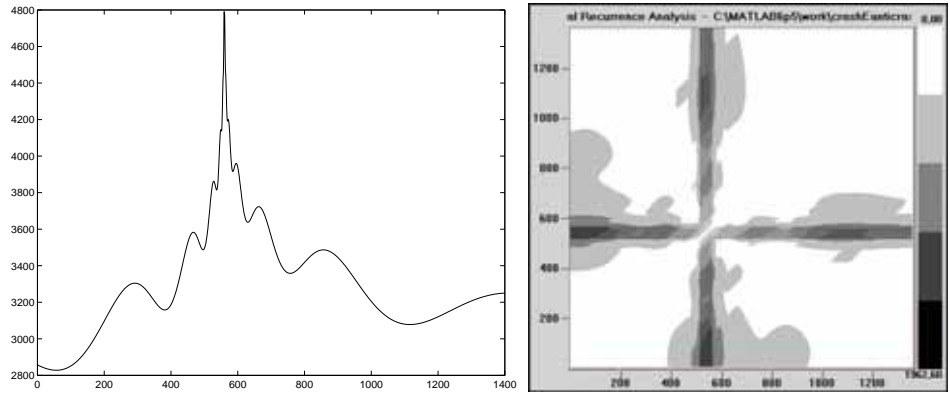

Figure 7. (RHS) the RP of a simulated phase transition of a signal (LHS) following the law (1.1) before and after the critical event.

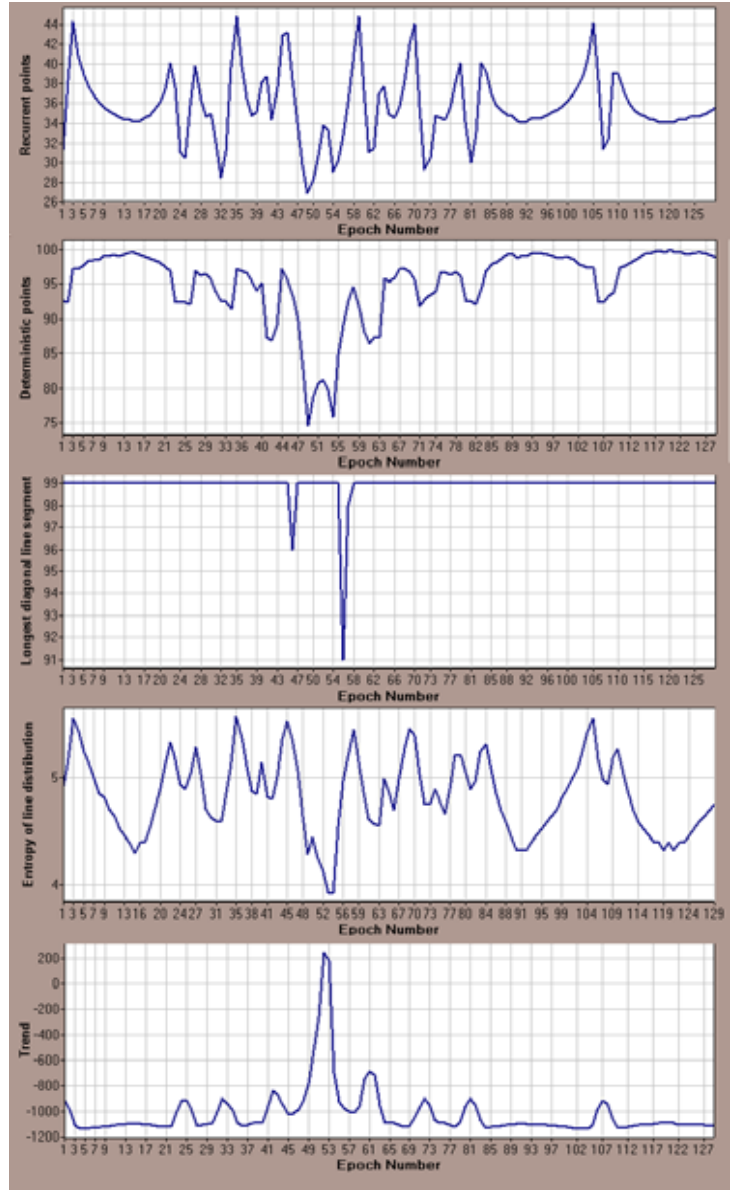

FIGURE 8. The values of the RQA variables for the simulated log-periodic signal (Fig 7) before and after the phase transition. Note that %REC, %DET, ENT and TREND take the lowest values in correspondence of phase transition period, see epochs 49 and 51, while MAXLINE takes the lowest values in epochs 42 e 57, respectively before and at the beginning of the transition period.

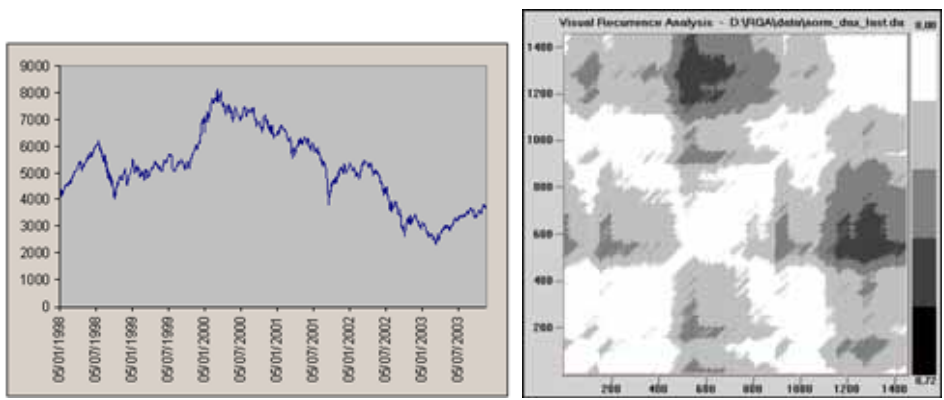

FIGURE 9. (LHS) daily closing price for DAX from Jan. 05, 1998 to Nov. 23, 2003; (RHS) RP of DAX from Jan. 05, 1998 to Nov. 23, 2003. The grey band with an 'L' shape almost centered around the point of coordinates (530,200) in the RHS plot corresponds to the peak reached in March 2000. The dark grey band on the right centered in $x = 1300$ represents the difference between the bullish evolution at the beginning of 2000 and the bearish one in 2003

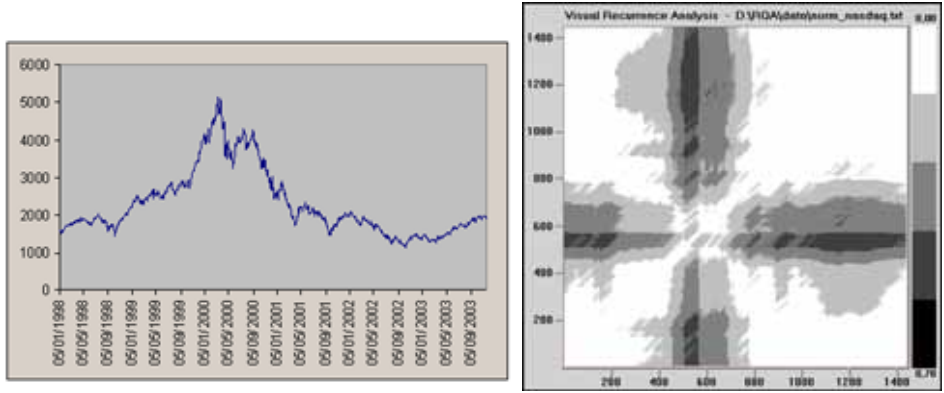

FIGURE 10. (LHS)daily closing price of NASDAQ from Jan. 05, 1998 to Nov. 21, 2003; (RHS) RP of NASDAQ from Jan. 05, 1998 to Nov. 21, 2003. The dark grey band delimited by horizontal coordinates $x = 504$ and $x = 566$, encircled by a grey area delimited by horizontal coordinates $x = 452$ and $x = 690$, is the 'image' of the crash of April 2000. It is a 'strong event' but afterwards the normal regime is restored

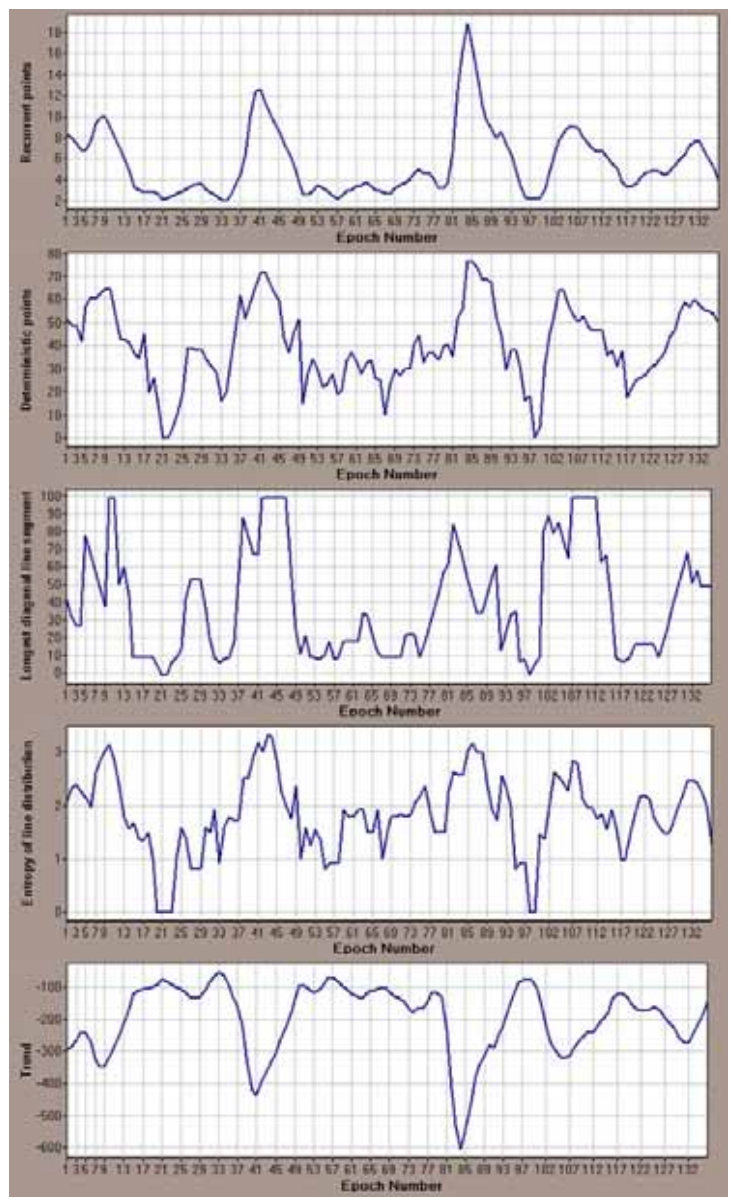

Figure 11. The values of the RQA variables for the daily closing price of DAX in 100 days epochs shifted by 10 days. It is worth to note that the highest absolute values are in epochs 38/44 and 82/87, corresponding respectively to the periods of the highest DAX peaks (July 1999 and March 2000) and to the deepest 'valleys' (March, 2001 and Sept., 2001) reached by DAX in the time span.

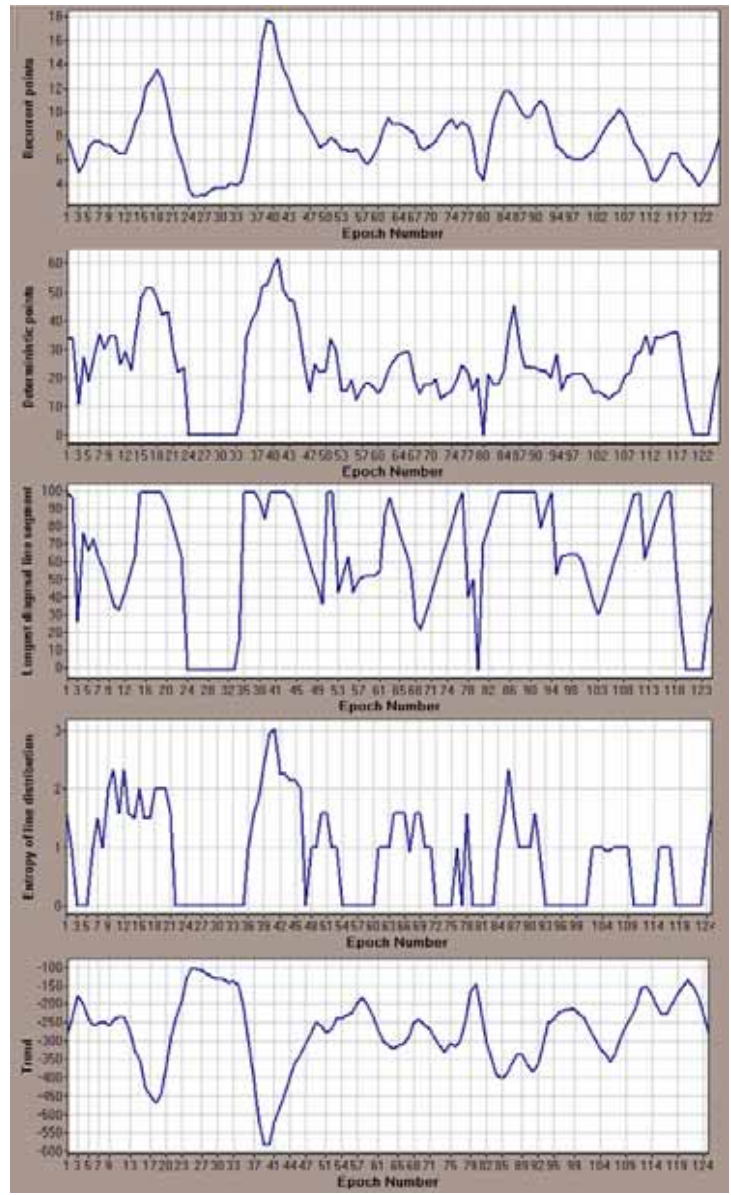

FIGURE 12. The values of the RQA variables for the daily closing price of NASDAQ in 100 days epochs shifted by 10 days. The highest absolute values are reached in epochs between 37 and 43, also corresponding to the period from June 10, 1999 to Jan. 28, 2000, when the bubble is growing.

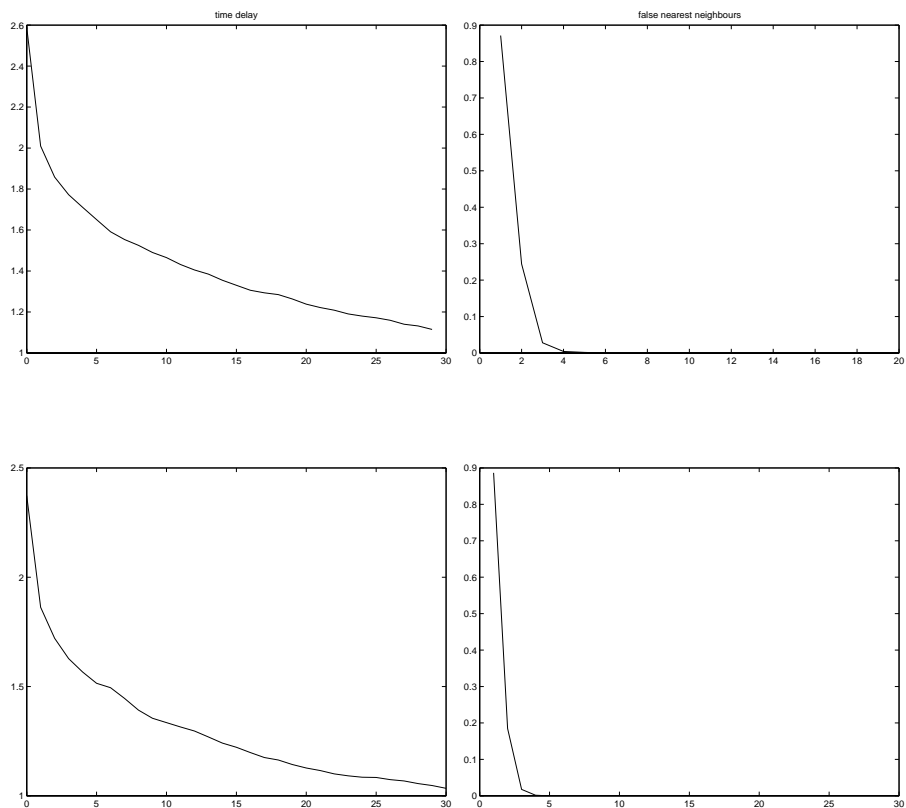

FIGURE 13. The two figures up are of DAX, while the figures down are of NASDAQ. (LHS) (up and down) output by TISEAN of the function 'mutual': on the $x$ axis there is the time delay, while on the $y$ axis the corresponding values of the mutual information. (RHS) Output by TISEAN of the function 'false_nearest', on the $x$ axis there is the embedding dimension while on the $y$ axis the relative percentage of fnn

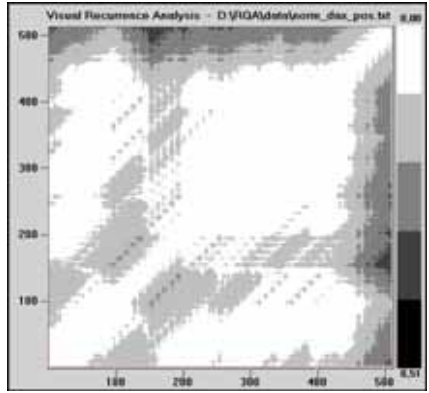

FIGURE 14. RP of DAX from Jan. 05, 1998 to March 08, 2000, corresponding to the bullish period. The large white area in the upper right corner is a stationarity period; instead, on the borders, the grey bands becoming darker indicate a drift.

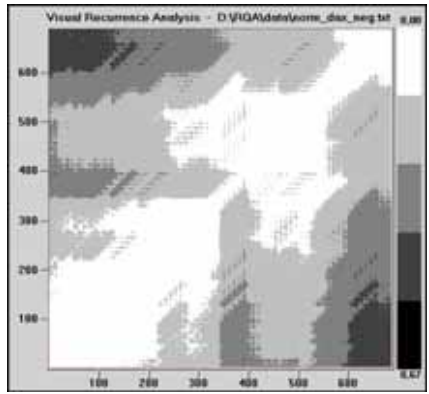

FIGURE 15. RP of the daily closing price of DAX from March 08, 2000 to Jan. 23, 2003. It is the RP of the bearish period; in the first 350 days the price has a soft negative trend that corresponds to a white area and a fading grey. The grey area delimited between coordinates $x = 350/400$ corresponds to a deep minimum in DAX on Sept. 21, 2001. After this minimum the price rises a little bit but the negative trend leads to an 'arrow' shape, i.e. the white area on the upper right corner centered in (450,500).

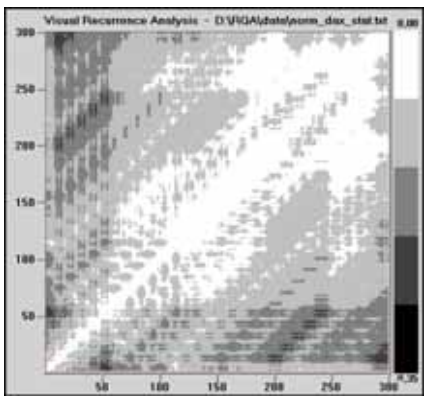

FIGURE 16. RP of Dax from July 15, 1998 to Nov. 29, 1999. The RP looks quite homogeneous.

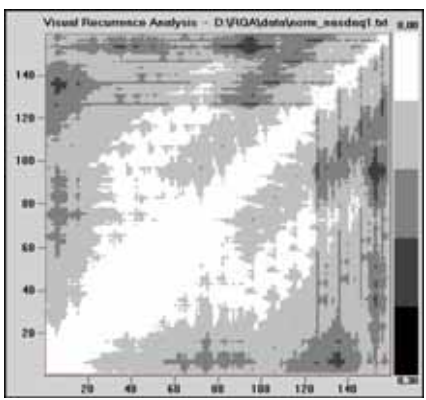

FIGURE 17. RP of the NASDAQ subseries (I) starting in Jan., 1998 and ending in Oct., 1998; the dark grey small area near $x = 130$ is the maximum NASDAQ price around 2000, in July 1998; except this the remaining part of the signal can be assumed stationary because the RP looks homogeneous.

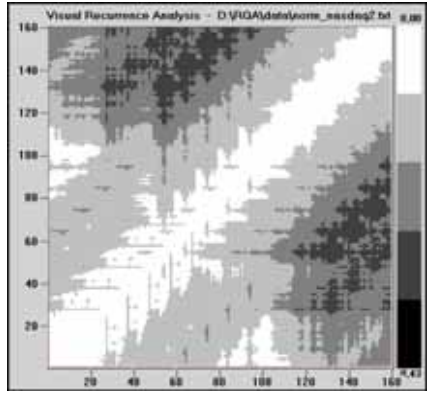

Figure 18. RP of the NASDAQ subseries (II) from June, 1998 to Feb., 1999; this RP is quite homogeneous except for some isolated not recurrent points aligned along the main diagonal

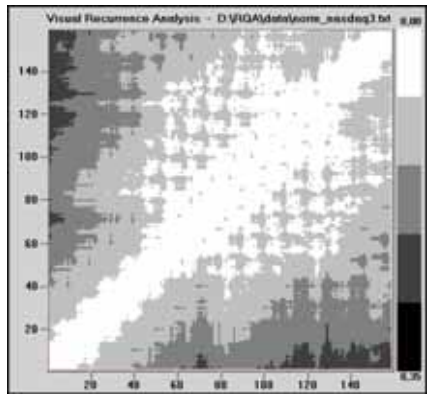

Figure 19. RP of the NASDAQ sub series (III) between Oct., 1998 and July, 1999

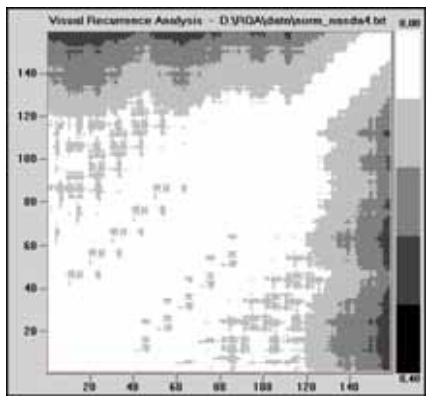

FIGURE 20. RP of the NASDAQ subseries (IV) from Jan., 1999 to Dec., 1999. This RP assumes the characteristic 'arrow' shape typical of a strong trend: it starts to be significant in the middle of Oct. 1999. It has to be underlined that this period does not include the crash time, but stops before the bubble bursts. It is worth to note also a similarity between this RP and the RP in figure 6, i.e. the simulation of the log-periodic function (1.1).

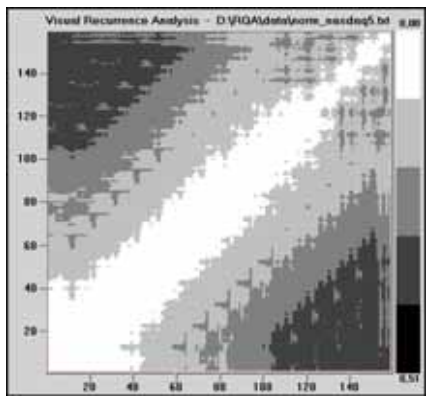

FIGURE 21. RP of the NASDAQ subseries (V) from Jul., 1999 to May, 2000; the grey area starts when the bubble starts; however in this RP, the bubble is not so evident because it is not possible to see the change between normal and critical regimes as in Fig 20.

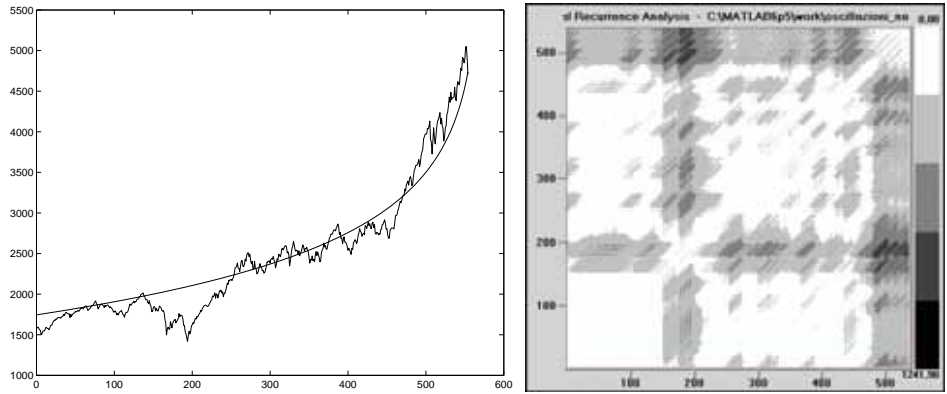

FIGURE 22. (LHS) the Nasdaq before the crash de-trended by function 3.1 (the smooth line). (RHS) its RP. It was expected to see signs of the log periodicity but due to the presence of noise, we are not allowed to affirm so.

TABLE 1. Embedding parameters for simulated data

| Embedding Parameters | time delay d | emb. dimension m |
|---|---|---|
| Brownian Motion | 2 | 3 |
| Sine | 3 | 2 |
| White Noise | 1 | 2 |
| Trend | 10 | 6 |
| Log-Period. | 7 | 3 |
| Function 1.1 | 11 | 5 |

TABLE 2. RQA of Brownian Motion, Sine, White Noise

| Series | Brownian motion | Sine | White Noise |
|---|---|---|---|
| size | 1000 | 1000 | 1000 |
| %REC | 18 | 15 | 0.6 |
| %DET | 62 | 69 | $\sim 0$ |
| MAXLINE | 397 | 997 | 0 |
| ENT | 4.3 | 3 | $\sim 0$ |
| TREND(units/1000points) | -16 | -0.8 | $\sim 0$ |

TABLE 3. RQA of simulated signals: Trend, Log-periodicity as in Function 3.2, Function 1.1

| Subseries | Trend | Log period. | function 1.1 |
|---|---|---|---|
| size | 1000 | 1000 | 400 |
| %REC | 37.9 | 19.2 | 39 |
| %DET | $\sim 100$ | 89.8 | 99 |
| MAXLINE | 997 | 987 | 383 |
| ENT | 8.5 | 6.0 | 7.3 |
| TREND(units/1000points) | -113.8 | -3.9 | -245 |

TABLE 4. RQA of Dax on the 3 studied periods: bullish corresponding to Jan. 05, 1998 and March 08, 2000; bearish: March 08, 2000 and Jan. 23, 2003; stationary: Jul. 28, 1998 and Nov. 29, 1999. Note that the lowest values are in the period assumed to be stationary

| Subseries | Bullish period | Bearish period | Stationary period |
|---|---|---|---|
| size | 552 | 728 | 341 |
| %REC | 48 | 54 | 15.3 |
| %DET | 12.8 | 9.3 | 2.8 |
| MAXLINE | 511 | 687 | 141 |
| ENT | 5.5 | 5.7 | 1.5 |
| TREND(units/1000points) | -38.8 | -47.1 | -31.3 |

TABLE 5. RQA of NASDAQ on 5 the subseries studied of 200 days. The subseries (I) starts in Jan. 1998 and ends in Oct. 1998; the subseries (II) starts in June 1998 and ends in Feb. 1999; the subseries (III) starts in Oct. 1998 and ends in July 1999; the subseries (IV) starts in Feb. 1999 and ends in Dec. 1999; the subseries (V) starts in July 1999 and ends in May 2000. Note that the subseries IV takes the highest values

| Subseries | I | II | III | IV | V |
|---|---|---|---|---|---|
| %REC | 6.075 | 9.141 | 5.513 | 17.146 | 9.246 |
| %DET | 35.980 | 36.119 | 29.079 | 45.018 | 54.511 |
| MAXLINE | 125 | 158 | 83 | 179 | 166 |
| ENT | 2.522 | 3.547 | 2.585 | 4.054 | 3.301 |
| TREND(units/1000points) | -105.125 | -155.824 | -97.808 | -273.775 | -138.501 |